\newcommand{\CLASSINPUTinnersidemargin}{18mm}
\newcommand{\CLASSINPUToutersidemargin}{12mm}
\newcommand{\CLASSINPUTtoptextmargin}{20mm}
\newcommand{\CLASSINPUTbottomtextmargin}{25mm}
\documentclass[10pt,conference,a4paper]{IEEEtran}

\usepackage[utf8]{inputenc}

\usepackage{times}

\usepackage{graphicx}
\usepackage{subfigure}
\DeclareGraphicsExtensions{.png,.eps,.ps,.pdf}

\usepackage{url}
\usepackage{nameref,hyperref}

\usepackage[english]{babel}
\usepackage{comment}

\begin{document}

\title{ECYSAP EYE: From Cyber Situational Awareness to Mission-Centric Decision Support\\
for Enhanced Cyberspace Operations}

\newcommand{\orcid}[1]{\href{https://orcid.org/#1}{\includegraphics[width=0.32cm]{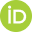}}}
\author{
    \IEEEauthorblockN{
    Pantaleone Nespoli\IEEEauthorrefmark{1}\orcid{0000-0002-4041-1205},
    Daniel Díaz-López\IEEEauthorrefmark{1}\orcid{0000-0001-7244-2631}, 
    Sergio López Bernal\IEEEauthorrefmark{1}\orcid{0000-0003-1869-1965},\\
    Francisco Oliva Bermejo\IEEEauthorrefmark{2}\orcid{0000-0000-0000-0000},
    Pedro González Megías\IEEEauthorrefmark{2}\orcid{0000-0000-0000-0000},
    Jorge Maestre Vidal\IEEEauthorrefmark{2}\orcid{0000-0002-4131-5100},\\
    Víctor Sobrino García\IEEEauthorrefmark{3}\orcid{0009-0001-2803-6684},
    Gregorio Martínez Pérez\IEEEauthorrefmark{1}\orcid{0000-0001-5532-6604}
    }
        
    \IEEEauthorblockA{\IEEEauthorrefmark{1}\textit{Departamento de Ingeniería de la Información y las Comunicaciones, Universidad de Murcia}, 30100 Murcia, Spain
    \\\{pantaleone.nespoli, danielorlando.diaz, slopez, gregorio\}@um.es}

    \IEEEauthorblockA{\IEEEauthorrefmark{2}\textit{Indra Sistemas S.A.}, 
    Av. de Bruselas 35, 28108, Madrid, Spain
    \\\{fjolivab, pgmegias, jmaestre\}@indra.es}

    \IEEEauthorblockA{\IEEEauthorrefmark{3}\textit{Universidad Politécnica de Madrid}, 
    Campus de Montegancedo, 28660 Boadilla del Monte, Madrid, Spain
    \\victor.sobrino@alumnos.upm.es}
    
}

\maketitle

\begin{abstract} 
Operational organizations increasingly require Cyber Situational Awareness (CySA) capabilities that go beyond isolated technical alerts, providing mission-relevant artefacts that can be embedded into heterogeneous toolchains and cyber security or cyber defense processes. ECYSAP EYE addresses this need through an adoption-oriented System-of-Systems (SoS) architecture centered on seven groups of mission-focused artefacts: the Recognized Cyberspace Picture (RCyP), Cyber Situational Reports (CySRs), the What-If Analysis Report (WIAR), Option Recommendations (OPRE), an operator Dashboard/HMI (DSH), Action Enforcement (AE), and After-Action Reports (AAR). The ECYSAP EYE architecture structures the transition from perception (full-spectrum RCyP views), to decision-oriented reasoning (WIAR/CySRs/OPRE), and to operational execution and learning (DSH/AE/AAR), with explicit integration surfaces that support incremental deployment and validation. This paper presents this innovative project from a technology transfer perspective, summarizing the updated architecture, the functional role of seven groups of artefacts, and the expected impact of cyber situations on the decision-making process in the context of a mission planning and execution.
\end{abstract}

\begin{IEEEkeywords}
Cyber situational awareness, mission engineering, decision support, course of action, system-of-systems
\end{IEEEkeywords}

{\bf Contribution type:}  {\it Transference track}

\section{Introduction}

Cyber operations have evolved from isolated technical incidents to sustained and adaptive campaigns capable of degrading critical services while shaping the outcome of military operations in the cyber domain. In the cyber operations environment, cyber effects rarely remain confined to the network perimeter. In fact, disruptions in Communications and Information Services (CIS), compromise of core systems, or manipulation of information flows can propagate to mission assets, degrade combat functions, and ultimately jeopardize mission success. As a result, effective cyber operations require the ability to contextualize cyber actions, anticipate their resulting effects, and support decision-making under time pressure, uncertainty, and operational constraints~\cite{MissionKott2017}; which also paves the way for their integration with other domains.

Situational Awareness (SA) is typically described as a three-stage process: (i) \textit{perception} of relevant elements, (ii) \textit{comprehension} of their meaning, and (iii) \textit{projection} of their future status to support timely actions in complex environments~\cite{endsley2015final}. In operational cyberspace, the EU and NATO refer to this capability as Cyber Situational Awareness (CY-SA), defined as the human ability to gain and maintain current and predictive shared knowledge of cyberspace to support decision-making. Achieving this requires supportive technological solutions capable of producing, maintaining, and disseminating information on friendly and adversarial forces, enabling cyber-focused operational assessment. Beyond data collection (e.g. threat intelligence, anomaly detection, configuration analysis), CySA is supported by tools that integrate heterogeneous signals into a coherent picture, reasoning about mission dependencies, and issuing actionable responses aligned with commander's intent. However, CySA's acquisition remains challenging due to the distributed, incomplete, and dynamic nature of cyber data, the adaptive behavior of adversaries, and the need to balance cyber risks with mission priorities and resource constraints~\cite{VisualizationJiang2022}.

In  light of the above, ECYSAP EYE project aims to advance prior ECYSAP work by strengthening the mission-centric view of cyber risk and by operationalizing decision support through a portfolio of consumable outputs (a Recognized Cyberspace Picture (RCyP), Cyber Situational Reports (CySRs), What-If Analysis Report (WIAR), Option Recommendations (OPRE), Dashboard (DSH), Action Enforcement (AE), and After-Action Reports (AAR)) supported by a modular system-of-systems architecture and an iterative engineering pathway oriented to TRL 6-7 deployment~\cite{CSAMontero2025}.

This paper focuses on ECYSAP EYE project's technology transfer value, i.e., how its concepts and products can be adopted by cyber defence practitioners and integrated into real combat ecosystems.
The article provides a concise transfer-oriented overview of ECYSAP EYE, highlighting:
\begin{itemize}
    \item A high-level SoS architecture of ECYSAP-EYE, focusing on seven operational artefacts that structure the architecture beyond a single shared picture.
    \item A concise functional description of the core modules/artefacts and how they support end-to-end mission-centric decision cycles from perception to reasoning and operational action/learning.
    \item A transfer discussion grounded on modularity, completeness and integration with other defence systems.
\end{itemize}

\section{A EU Cyber Situational Awareness Platform}\label{sec:ecysap_to_eye}

European collaboration for military-focused cyber situational awareness initiated with early concepts and specifications sustained by common requirements from multiple Member States (MS), under the coordination of the European Defence Agency (EDA). Among the results that preceded ECYSAP, in 2019 the Cyber Defence Situation Awareness Package Rapid Research Prototype (CySAP-RRP, TRL 4) delivered and proved feasibility in joint CySA related developments, covering dynamic risk assessment and mission impact analysis, considering both mission's attached Communications and Information Services (Technical) and their relevance for the commander's intent (mission)~\cite{EDA_CySAP_RRP_2020}.

The project ECYSAP (2020-2024) expanded CySAP-RRP into a broader platform for real-time CySA, adding rapid response functionalities and expanding its decision support functions to satisfy the needs of a wider community of EU MS and their cyber forces. ECYSAP achieved TRL 6-7 and proved cross-national interoperavility under NATO's Federated Mission Network (FMN) standards, being able to operate on CLASS/UNCLASS networks. Early results were shown in large military exercises, in support of tactical and operational planning teams, which motivated further transfer and harmonization activities~\cite{EC_ECYSAP}.

Building on these foundations, a new development spiral refereed to as ECYSAP EYE (2025-2028) is expanding past capabilities by introducing novel: (i) mission-centric products for decision support, (ii) engineering mechanisms to explore risks/opportunities and adversarial reactions, and (iii) added system integration and deployment readiness aligned with doctrine and training needs.

\begin{figure*}[th!]
  \centering
  \includegraphics[width=0.52\textwidth]{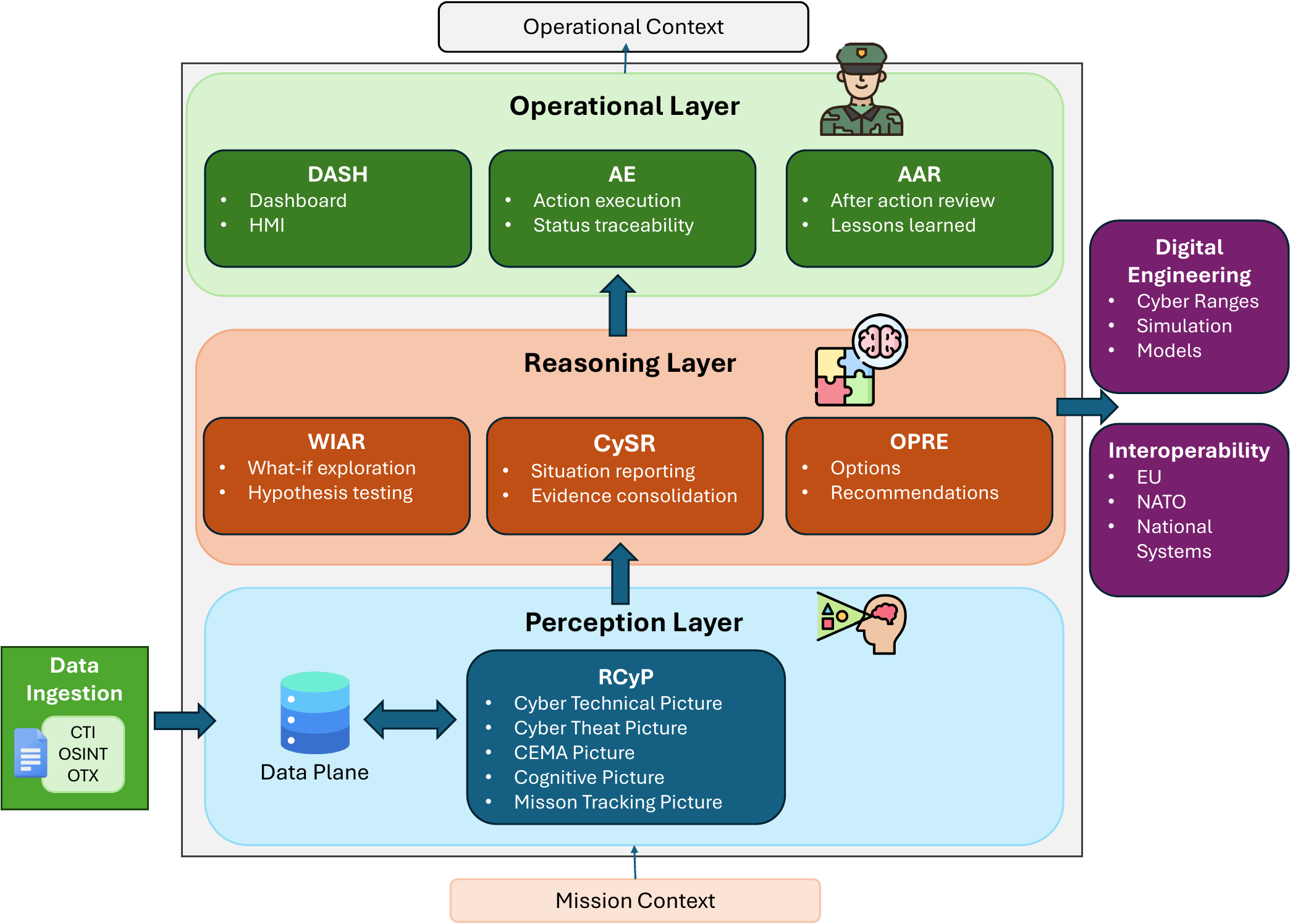}
  \caption{High-level System-of-Systems (SoS) architecture vision for ECYSAP EYE}
  \label{fig:sos_arch}
\end{figure*}

\section{The ECYSAP EYE's SoS Architecture}\label{sec:sos_arch}

A central objective of ECYSAP EYE is to provide mission-centric CySA through a \textit{modular SoS architecture}. This enables: (i) integration of heterogeneous data sources and external services, (ii) deployment across multiple nodes with differentiated responsibilities, and (iii) progressive adoption through incremental capabilities onboarding. Particularly, this architectural approach is relevant for technology transfer, as it reduces integration friction and supports interoperability with existing national/EU/NATO ecosystems.

The operational value is delivered through a set of \emph{artefacts} elaborated by ECYSAP EYE's operators and consumed by mission stakeholders. Figure~\ref{fig:sos_arch} summarizes the updated high-level architecture as a layered stack that connects (i) operational decision-making, (ii) reasoning and reporting, and (iii) data-driven perception through a full-spectrum RCyP. 

At the bottom, the data ingestion component integrates external and contextual sources (e.g. Cyber Threat Intelligence (CTI), Open Source Intelligence (OSINT) and Open Threat Exchange (OTX)) that feed a full-fledged data plane. The latter provides the foundational services required to operationalize ECYSAP EYE, such as normalization, persistence, correlation, and traceability to obtain curated information products. On top of this base, the \textbf{RCyP service} produces a \emph{full-spectrum RCyP} comprising complementary pictures that jointly support mission-centric comprehension. These are tailored for answering each echelon needs for information.

In the middle layer, the \textbf{Reasoning layer} transforms full-spectrum RCyP views and \textbf{Mission Context} into decision-oriented artefacts. Specifically, \textbf{WIAR} supports structured what-if exploration and reasoning under uncertainty. Then, \textbf{CySR} provides situation reporting that consolidates evidence into mission-relevant narratives. Last but not least, \textbf{OPRE} generates option/recommendation outputs that support response.

Finally, the \textbf{Operational layer} exposes these artefacts to mission workflows. \textbf{DASH} (dashboard/HMI) acts as primary consumption interface, enabling shared visibility and coordination. Moreover, \textbf{AE} captures the execution and enforcement of selected actions, while \textbf{AAR} consolidates outcomes for learning and iterative improvement, closing the feedback loop of observation, decision, actuation and evaluation.

In addition to the three-layer stack, Figure~\ref{fig:sos_arch} highlights two external enablers that support operational deployment and technology transfer across heterogeneous environments. Specifically, Digital Engineering provides integration points with cyber ranges, simulation, and modeling toolchains, enabling controlled experimentation, training, and pre-deployment validation of the artefacts before operational rollout. In parallel, the Interoperability module captures the interfaces required to exchange information products with external military systems, ensuring that RCyP and downstream artefacts can be integrated into joint workflows and multi-stakeholder command structures rather than confined to a standalone platform.

\section{Core Modules and Functional Workflow}\label{sec:core_artifacts}

To align with operational adoption, ECYSAP EYE is best understood as a pipeline of \emph{core modules} that generate and consume artefacts across the mission decision cycle. Each module is designed to be integrated independently while preserving an end-to-end flow from observations to actuation and learning. While ingestion and data plane provide its technical substrate (collection, normalization, persistence, correlation, and traceability), the operational value and transfer focus lie in the utility of artefacts for mission stakeholders.

First, the RCyP is the artefact that provides a commmon and authoritative representation of the cyber situation, intended to be shared across echelons and that feeds the Common Operational Picture (COP) from the cyber domain perspective. RCyP is composed of complementary views (cyber technical, CEMA, Cyber Threat, Cognitive, and Mission Tracking), allowing human operators to maintain a consistent baseline that bridges cyber evidence to mission-relevant context. Functionally, RCyP acts as the primary reference for both reasoning outputs and operator-facing interaction. 

Then, WIAR operationalizes reasoning through structured exploration of hypotheses and \textit{what-if scenarios}, using RCyP to evaluate plausible evolutions and their consequences, providing the analysis layer that improves adaptive planning and CoA development. WIAR bridges perception with decision-ready understanding and projection under operational friction.

In addition, CySR consolidates relevant RCyP and reasoning outcomes into situation reporting  suitable for operational briefings and coordination. This reduces the need for human intervention when writing.

Also, OPRE produces option/recommendation outputs by linking assessed potential risks and opportunities to feasible operational responses derived from RCyP and WIAR outcomes. That is, this component supports prioritization and trade-offs (e.g. mission impact versus mitigation cost or constraints) and provides structured inputs for high-level decisions.

At a higher level, DSH can be described as the primary human operator-facing interface for consuming and coordinating the ECYSAP EYE's artefacts. Concretely, it supports shared visibility, role-based interaction, and integration into existing command chain workflows, acting as the practical ``entry point'' and awareness-ready module for operational use.

Moreover, AE captures the operationalization of selected decisions by supporting and tracking the enforcement of actions. Specifically, AE closes the gap between recommendations and real operational actions, providing status and traceability so that outcomes can be related back to the decisions and the underlying evidence and services.

Finally, AAR closes the loop by consolidating outcomes, and fostering the lessons identified and learned process while enabling operational assessment. AAR links observed effects (leveraging AE and the constantly evolving RCyP) to the decision process (i.e. WIAR and OPRE), supporting accountability, training, and iterative improvement of both procedures and configurations. This assessment artefact is particularly relevant for technology transfer, where trust-building and institutionalization typically require documented validation cycles to support potentially disruptive decisions.

\begin{table*}[th!]
\caption{Main ECYSAP-EYE's artefacts and transfer value (summary).}
\label{tab:products_transfer}
\tiny
\begin{tabular}{p{3cm} p{3.3cm} p{4.5cm} p{4.5cm}}
\hline
\textbf{Artifact/Service} & \textbf{Primary user} & \textbf{What it provides} & \textbf{Transfer value} \\
\hline \hline
\textbf{Recognized Cyberspace Picture (RCyP)} & Joint Operation Center (JOC)/external systems & Full-spectrum cyber picture (Cyber Technical, CEMA, Cyber Threat, Cognitive, Mission Tracking views) & Shared landscape across systems and domains; improves coordination and interoperability-ready consumption \\ \hline
\textbf{Cyber Situational Reports (CySRs)} & Command staff/operators & Reporting artifact consolidating evidence into mission-relevant narratives & Standardizes communication; improves comprehensive and coordinated conduction of operation \\ \hline
\textbf{What-If Analysis Report (WIAR)} & Analysts/staff & Structured what-if exploration (assumptions, hypotheses, projections, consequences) & Supports explainable reasoning; facilitates scenario-based validation and trust building; supports planning \\ \hline
\textbf{Option Recommendations (OPRE)} & Commander/planners & Options and recommendations linked to assessed risks/opportunities & Enables prioritization and traceable decision justification; supports repeatable procedures; enables alternate CoA development \\ \hline
\textbf{Dashboard (DSH)} & Operators/Commander & Operator-facing interface to consume RCyP, CySRs, WIAR, OPRE & Embeds artefacts into real workflows; lowers adoption barriers via usable operational interface \\ \hline
\textbf{Action Enforcement (AE)} & Response teams/operators & Enforcement tracking of selected actions and status/outcomes & Bridges decisions to operational effects; enables traceability and integration with toolchains; provides cost-effectiveness analysis \\ \hline
\textbf{After-Action Reports (AAR)} & Commander/Auditors/QA & Post-operation assessment: outcomes, lessons learned, feedback & Closes the learning loop; supports institutionalization, training, and continuous improvement \\ \hline
\hline
\end{tabular}%
\end{table*}


A key objective of ECYSAP EYE is to maximize operational adoption by reducing integration friction and enabling progressive onboarding of capabilities. From a transfer perspective, the relevant question is not only \textit{what} the system provides, but \textit{how} its seven groups of artefacts can be embedded into real ecosystems where tooling, data availability, military doctrine, and operational responsibilities differ across organizations (and potentially across nations). In addition to adopting interoperability standards like FMN, the SoS design in Figure~\ref{fig:sos_arch} supports this ambitious goal by structuring a clear transition from perception to reasoning to operational execution, with explicit integration surfaces that enable incremental deployment and validation across heterogeneous environments

\subsection{Transfer-Oriented Adoption: Integration Readiness}
The proposed architecture enables stakeholders to adopt the artefact stack progressively. At the perception level, the RCyP provides a shared, full-spectrum baseline  representation of the operational cyber landscape that can be exposed to both operators and external systems. On top of this baseline, reasoning artefacts (i.e., CySRs, WIAR, OPRE) can be introduced incrementally to support structured planning based on the information provided by mechanisms as what-if exploration, situation reporting, and options/recommendations, respectively. From an operational perspective, WIAR enables the preparation of alternative ways of action for contingency plans before missions. The RCyP provides the landscape status and configuration to the user. Once an event occurs, the OPRE will compute and provide response options in the cyber domain to officers for their approval, while the AE/AAR will monitor the event and the decision-making process to provide effect assessment and analysis to improve the effectiveness of subsequent recommendations and actions.

On the consumption side, DSH offers a stable integration point for command workflows by providing role-oriented access to the artefacts, while interoperability interfaces facilitate coupling with external EU/NATO/national systems where required. In addition, digital-engineering connectors (cyber ranges, simulation, models) support controlled testing and training, which are often prerequisites for operational rollout and institutionalization. 

The proposed architecture ensures cyber and operational domains are recognizable at both tactical and operational levels of war by first integrating information into mission and operational pictures, while providing commanders with the necessary tools that support their cyber operations by facilitating common understanding of the situation, and contributing to a more effective employment of their cyber forces. 

\subsection{Operator-Facing Outputs}
Table~\ref{tab:products_transfer} summarizes the seven artefacts and their primary transfer value. The emphasis is on delivering operator-ready artefacts that support mission-centric decision cycles, that is, a shared baseline picture (RCyP), decision-oriented reasoning products (CySRs, WIAR, OPRE), and operational loop closure (DSH, AE, AAR). 

In particular, RCyP operationalizes shared situational awareness by providing synchronized and traceable views (Cyber Technical, CEMA, Cyber Threat, Cognitive, and Mission Tracking pictures) that can be consumed consistently across teams and, when needed, external systems. In this way, RCyP is able to reduce ambiguities in joint decision-making and supports coordination under time pressure, fostering a unique and easily interpretable common operational picture that successfully integrates the cyber and cognitive domains.

Similarly, the reasoning artefacts are intended to produce interpretable outputs. Particularly, i) WIA captures structured what-if exploration grounded in mission context, enabling pre-planned response options, ii) CySRs consolidate evidence into mission-relevant narratives suitable for briefings and coordination, and iii) OPRE links assessed risks/opportunities to feasible response options and recommendations, expediting the development of CoAs and linking WIA planning with OPRE conduction. These artefacts are consumed through DSH, which acts as the primary operator interface and supports the transition from perception (RCyP) to comprehension and projection (WIAR/CySRs/OPRE), enabling timely decision-making during mission planning and execution.

Finally, operationalization requires traceability from decisions to effects and back to improvement. AE incorporates effect-based analysis of operations based on the recorded status and outcomes of the entire process, while AAR consolidates evaluation, operational assessment, and lessons learning. This explicit closure is particularly relevant for technology transfer, as operator trust and procedural alignment are typically established through iterative validation cycles rather than one-off integration efforts.

\subsection{Expected Impact}
By aligning cyber evidence with mission context and delivering a TRL 6-7 with an SoS architecture centered on operational artefacts, ECYSAP EYE is positioned to strengthen decision-making and coordination across mission planning and execution. The explicit inclusion of interoperability and digital-engineering integration supports transfer into operational environments, where evaluation, training, and interconnection with existing systems are prerequisites for adoption. In particular, combining a shared RCyP baseline with structured reasoning and action/learning artefacts (WIAR/CySRs/OPRE, AE/AAR) can reduce the gap between technical cyber assessments and mission-level decisions, enabling faster coordination among stakeholders and more consistent justification of response options.

\section{Conclusion}\label{sec:conclusion}
This paper presented ECYSAP EYE from a technology-transfer perspective, emphasizing the project’s artifact-centric approach and its layered SoS architecture that connects perception (RCyP), decision-oriented reasoning (WIAR, CySRs, OPRE), and operational execution (DSH, AE, AAR). The resulting decomposition provides a practical path from cyber evidence and mission context to mission-centric decision support while preserving deployment flexibility and integration readiness. It is important to remark that the project remains under development in close connection with the Ministries of Defense (MoDs) involved, with ongoing integration and validation activities aimed at operational adoption.

\section*{Acknowledgment}
\tiny EU \normalsize This work has been co-funded by the European Union (EDF program; project ECYSAP EYE). Views and opinions expressed are however those of the author(s) only and do not necessarily reflect those of the European Union or the European Defence Fund. Neither the European Union nor the granting authority can be held responsible for them.\\

\bibliographystyle{IEEEtran}
\bibliography{bibliography.bib}{}

\end{document}